\documentstyle[11pt,aaspp]{article}

\begin{document}
\title{Assessing the Accuracy of Masses and Spatial Correlations of Galaxy
Groups}
\author{James J. Frederic}
\affil{Department of Physics, MIT 6--110, Cambridge, MA 02139}

\begin{abstract}
Two algorithms for the identification of galaxy groups from redshift surveys
are tested by application to simulated data derived from N-body simulation.
The accuracy of the membership
assignments by these algorithms is studied in a companion to this paper
(Frederic 1994).
Here we evaluate the accuracy of group mass estimates and the group-group
correlation function.

We find a strong bias to low values in the virial mass estimates
of groups identified using the algorithm of Nolthenius \& White (1987).
The Huchra \& Geller (1982) algorithm gives virial
mass estimates which are correct on average.
These two algorithms result in group catalogs with similar two-point
correlations.

We find that groups in a CDM model have excessively large
mass to light ratios even when the group richness distribution
agrees with observations.  We also find that our
CDM groups are more strongly correlated than individual
halos (galaxies),
unlike the groups in the CfA redshift survey extension.
\end{abstract}


\section{Introduction}
Dynamical studies of groups of galaxies are important for probing
the galaxy and the mass distributions in the universe over
two orders of magnitude in mass, between the small scales
probed by the internal velocities of individual galaxies (up to
about $10^{12}M_{\odot}$)
and the larger scales relevant to rich clusters of galaxies (over
$10^{14}M_{\odot}$).
Previous workers (see Paper I for references),
many using different and often subjective criteria
for defining groups, have identified groups, estimated their masses
and other internal properties, and characterized their clustering properties.
A general difficulty in this work has been the determination of reliable error
estimates.

In this paper we are concerned with identifying galaxy groups and
measuring their internal properties (masses, mass to light ratios,
internal velocity dispersions, etc.) as well as their clustering properties.
We study two algorithms for identifying
galaxy groups from redshift surveys.  The first was introduced
by Huchra \& Geller (1982, hereafter HG82) and used by the same
authors in Geller \& Huchra (1983, hereafter GH83) and again by Ramella,
Geller \& Huchra (1989, hereafter RGH89).  The second was
proposed by Nolthenius \& White (1987, hereafter NW87) as
an improvement to the first.  We test these algorithms by applying them to
simulated redshift survey data obtained from an N-body
simulation performed by Gelb (1992).
A companion paper, Frederic (1994, hereafter Paper I), focuses on
the accuracy of the membership assignments for each group finding algorithm.
Here we study the accuracy of the group mass estimates and correlation
function.

We optimize the group finding algorithms by applying them
to the redshift survey catalogs constructed from our simulation.
Once the limitations of the group finding algorithms have been determined,
we can apply the optimized algorithms to real data.
We use the first $6^{\circ}$ declination slice of the Center for Astrophysics
(CfA)
redshift survey extension complete to $m_B=15.5$ (Huchra, J.P., {et al.}
1990)
The data were obtained electronically through
the Astronomical Data Center of the National Space Science Data
Center/World Data Center A for Rockets and Satellites at NASA Goddard Space
Flight Center.  We used the February
1992 version of the catalog, ADC catalog number 7144 (Huchra, J.P., {et al.}
1992).

Once a group catalog has been constructed and uncertainties in group
membership have been quantified,
group masses may be determined by application of the virial
theorem or similar techniques (Heisler, Tremaine, \& Bahcall, 1985).
Application of these methods to simulated groups of known mass allows
determination of
the uncertainty inherent in the dynamical mass estimates.

Paper I contains descriptions of the N-body experiment and DENMAX, the method
of galaxy identification in the simulation, as well as the method
by which magnitude limited redshift catalogs were generated from
the simulation.
An important consideration dealt with in Paper I
is the overmerging problem, common to high resolution N-body simulations
of collisionless matter, in which small clumps of mass merge into
extremely large clumps with almost no substructure.
Specific group finding algorithms were then discussed
and their accuracy studied.  In this paper, section 2 describes
the group finding algorithms.  Section 3 is concerned with
the internal properties
of groups, such as masses and mass to light ratios.  The group-group
correlation functions are studied in section 4.
Section 5 presents conclusions.  An appendix containing definitions
of various statistics follows.

\section{Group Finding Algorithms}
Groups are identified by both the HG82 algorithm and
by the NW87 algorithm.  Hereafter HG and NW will refer to the
algorithms and to the groups constructed using the algorithms.
Each operates on a
list of galaxy angular positions and redshifts.  Each pair of galaxies
which are separated by less than some specified amount in both
redshift and projected separation is considered linked.  Each distinct set
of mutually linked galaxies is a group.

The radial and transverse linking lengths for the HG algorithm are given by
\begin{equation}
D_L=D_0{\left[\int_{-\infty}^{M_{V}} \phi(M)dM \bigg/
\int_{-\infty}^{M_{\rm lim}} \phi(M)dM\right]}^{-1/3} \,,
\label{dl}
\end{equation}
\begin{equation}
V_L=V_0{\left[\int_{-\infty}^{M_{V}} \phi(M)dM \bigg/
\int_{-\infty}^{M_{\rm lim}} \phi(M)dM\right]}^{-1/3} \,,
\label{vl}
\end{equation}
where $M_{V}=m_{\rm lim}-25-5\log(V/H_0)$ is the absolute magnitude
of the brightest galaxy visible at a distance $V/H_0$ and $V$
is the mean velocity of the pair.  Similarly,
$M_{\rm lim}=m_{\rm lim}-25-5\log(V_F/H_0)$ is the absolute magnitude of
the brightest visible galaxy at a fiducial distance $V_F/H_0$.
$D_0$ and $V_0$ are the linking cutoffs at $V_F$, and $\phi(M)$ is
the galaxy absolute magnitude (luminosity) function.

The same quantities for the NW algorithm are
\begin{equation}
D_L=D_0 \left[ {\int_{-\infty}^{M_{V}} \phi(M)dM \over
\int_{-\infty}^{M_{\rm lim}} \phi(M)dM} \right]^{-1/2}
\left[{V_F \over V} \right]^{1/3} \,,
\label{nwdl}
\end{equation}
\begin{equation}
V_L=V_0+0.03(V-5000~{\rm km~s^{-1}}) \,.
\label{nwvl}
\end{equation}

In each algorithm, the linking lengths increase with redshift to account
for the declining selection function.  The primary difference between
the two algorithms is in the scaling of the velocity linking length,
which increases more slowly with redshift for the NW than for
the HG algorithm.
The algorithms are described in greater
detail by the original authors and in Paper I.

These algorithms are applied to simulated redshift surveys
constructed by two methods.  In the first, which we call our ``raw''
catalog, each halo formed
in the simulation is treated as a single galaxy.  To construct
our ``breakup'' catalog, halos which are too large as a result
of overmerging are split into several galaxies based on an assumed
mass to light ratio for clusters and the galaxy luminosity function.
In either case, simulated galaxy luminosities are selected
to fit the Schechter (1976) form of the luminosity function
as determined by de Lapparent, Geller, \& Huchra (1988) for the
first slice of the CfA survey extension.  The rank order of halo
circular velocities is preserved, so that the $n$th brightest
halo has the $n$th largest circular velocity.
The details of these procedures are given in Paper I.

We construct group catalogs using three variants
of each of the HG and NW algorithms, from ten (corresponding to
ten different observers) simulated redshift surveys.  Ten observers are used
to produce enough groups to give good statistics.  Of the three variants
on the group finding algorithms, one is the basic algorithm, where
groups are identified from an apparent magnitude limited redshift survey
using only the observationally available coordinates of redshift,
right ascension, and declination.  We call the result a catalog of Vm
groups, with V signifying the use of velocity (redshift) as the radial
coordinate and m indicating that the source galaxy list was apparent
magnitude limited.  In order to isolate the effects of peculiar velocities
on group identification we also construct our Rm catalog, in which
real distance rather than velocity is used as the radial coordinate.
In this case we use the same link ($D_L$, eqs. [\ref{dl}] and [\ref{nwdl}])
in both the radial and transverse directions.
Finally, we construct our RM catalogs, which are also based on true distances
but have as their source an absolute magnitude limited galaxy list.
The RM catalogs allow us to evaluate the effect of the magnitude limit
on group properties.  We construct RM catalogs using a fixed linking length.
To summarize, for each of ten observers in
two halo catalogs (breakup and raw), we
construct three types (Vm, Rm and RM) of catalogs based on two
grouping algorithms (HG and NW), for a total of 100 distinct
group catalogs. (RM catalogs do not distinguish between HG and NW.)
Unless explicitly noted, all galaxies and groups referred to are from
simulation; real groups from the CfA survey are denoted as such.

\section{Group Properties}
Observers of galaxy groups have been particularly concerned with group
mass to light ratios, under the assumption that the typical group
mass to light ratio is representative of the universe as a whole.
In particular, the median mass to light ratio of groups may be multiplied
by the luminosity density of the universe to estimate the mean mass
density.  The luminosity function used to illuminate our simulated
halos corresponds to a luminosity density of
$1.035 \times 10^8~h~L_\odot~{\rm Mpc}^{-3}$.  Thus, a universal
mass to light ratio of $1341h$ corresponds to $\Omega=1$.
We compute mass to light ratios for our simulated groups, but emphasize
that any results based on mass to light ratios must be interpreted
conservatively, as they are likely to be sensitive to the manner in
which our dark halos were illuminated.  For this reason we prefer
to consider the accuracy of our group mass estimates, obtained
with the standard virial theorem (Heisler et al. 1985), rather than mass
to light ratios.

In order to determine the accuracy of our virial theorem mass
estimates, we must compute the true masses of groups
in the simulated catalogs.  Our halo identification procedure, DENMAX,
associates particles with peaks in the evolved density field.
A further procedure removes those
particles which are not gravitationally bound to a halo.  The result
is a mass for each N-body halo which we denote $M_{\rm halo}$.
Because the virial theorem
is insensitive to mass which is not part of the system
for much of its evolution, we expect the virial masses of the groups
to measure the bound masses of the halos, plus any mass which may be bound
to the group as a whole but not to any individual halo.  In order to determine
how much mass might be lurking in our simulated groups, unattached to halos,
we counted the number of particles within one Abell radius ($1.5h^{-1}$ Mpc)
of the centers of the illuminated halos in our raw halo catalog.  We found
that 69\% of the mass in the simulation was within an Abell radius
of a halo, and that 54\% of the mass was bound to the halos.  Therefore,
at most 15\% of the mass may be in groups but not bound to any halo.  This
estimate
is certainly high, since it counts mass near all halos, not just halos
in groups.  And some fraction of this mass which is near but not bound to a
halo is probably not bound to any group, despite being within an Abell radius,
and therefore would not
be reflected in the virial masses.  So when computing the true mass of
a group, ignoring the mass which is not bound
to any illuminated halo will probably result in an undercounting of the true
mass by less than 10\%.

In order to determine the total true mass of a simulated group we must
know which of the galaxies fainter than the magnitude limit are to be
included in the group.
In the most general case, a given Vm or Rm group may
contain some members which are common to a larger RM group,
some other members which are equivalent to an entire RM group,
and still other members which are not in any RM group.
We choose to consider the true masses $M_{\rm true}$
of our Vm and Rm groups to be the sum of the
masses of the RM groups which overlap them in whole or in part.  That is,
we sum the masses of the members of each RM group, which is
complete down to an absolute magnitude $M_{B(0)}=-16.0$ (corresponding
to a halo of 6 particles), and assign those
masses to the Vm or Rm group which contains members of the RM group.
We also computed, instead of the total mass of the RM components, the mass
of the most massive RM component.  In many cases these masses are identical.
We prefer the total mass of the RM components to the largest mass component
because the total mass includes those small RM groups that are below the
magnitude limit in the Vm and Rm catalogs.  Considering only the largest
fraction of this mass would be ignoring mass which is near the Vm or Rm group
and has contributed to the gravitational binding of the group.

\subsection{Internal Properties of Simulated Groups}
In Paper I we studied the accuracy of the group finding algorithms
by comparing the memberships of redshift space groups to those of real space
groups.  It is possible, however, that accurate group memberships
are not essential for computing certain statistics such as the mean
or median of the distribution of a particular group property.
RGH89 argue in favor of this point, claiming in particular that
the median $M/L$ of their
most accurately
identified groups was essentially the same as the median $M/L$ for their
entire sample of groups, including those contaminated by interlopers.

Here we consider the accuracy of the group catalogs, as opposed
to the individual groups.  To do so we compare the distributions of various
group properties for our real space and our redshift space groups.
Tables 1 through 4 present statistics of the simulated Vm group catalogs.
They are the number of members $N$,
the true group mass $M_{\rm true}$,
the virial theorem mass estimate $M_{\rm VT}$, a measure of the
accuracy of the mass estimate $M_{\rm VT} / M_{\rm true}$, the mean harmonic
radius $r_h$,
the dispersions
in both velocity $\sigma_v$ and true distance $\sigma_r$, the crossing
time $t_c$, and the mass to light ratios $M_{\rm true}/L$
and $M_{\rm VT}/L$.
The true distance dispersion $\sigma_r$ and the crossing time $t_c$
are presented as an indication of the likelihood of group accuracy.
Small $\sigma_r$ indicates compactness in space and small $t_c$
means the groups have had time to virialize.
Equations defining all these quantities are given
in the appendix.  The tables give the mean and the
three quartiles of the distributions for each quantity.  Each entry
in the tables gives a mean and a standard deviation obtained
from our ten different observers.  This same information is given
for Rm groups in Tables 5 through 8, and for RM groups in
Tables 9 and 10.
Of course, not all of the above quantities are known for groups
selected from a real redshift survey.  Those that are are presented
for the HG and NW groups in the CfA data in Tables 11 and 12.

The distributions of many of these statistics differ systematically
between raw and breakup catalogs, between HG and NW grouping algorithms,
and between Vm and Rm groups.  When group identification is
performed in redshift space in the apparent magnitude limited (Vm) galaxy
catalogs, breakup groups tend to have higher true mass than do groups
in the raw catalogs.  Two effects are at work here.  One is that some
of the massive halos which were isolated and ungrouped in the raw catalog
become
rich, massive groups after breakup, thus raising the distribution
of true group masses.  The other is that due to the conservation
of luminosity in the breakup procedure, faint, low mass groups
in the raw catalogs disappear in the breakup case.
These effects also cause the breakup groups to have higher
median velocity dispersions and lower radii and crossing times than the
raw groups.  Group virial crossing times, proportional to $r_h / \sigma_v$,
are therefore lower for the breakup groups.  Since groups with
crossing times greater than the Hubble time cannot be virialized,
lower crossing times should generally result in more accurate virial
masses.  This explains
why the virial mass estimates are better for the HG breakup groups
than for the HG raw groups.

NW and HG groups differ predominantly in their
velocity dispersions and therefore in their virial masses.
Comparing raw HG to raw NW and breakup HG to breakup NW redshift space (Vm)
groups, we see the
primary difference between the two algorithms.  The smaller velocity
linking length used in the NW algorithm biases the group velocity
dispersions, and hence the virial masses, to low values.  The distributions
of true masses for the HG and NW groups are almost identical, indicating
that the two algorithms tend to pick out the same RM groups.  However,
by excluding the high velocity dispersion members of true groups, the
NW algorithm leads to low virial mass estimates.
As expected, this bias does not occur in the Rm catalogs.  In fact, in
real space the NW algorithm picks out groups with
slightly higher velocity dispersions than the Rm HG groups.

After breakup, the virial masses of the HG groups are larger
than before breakup, while the NW groups have smaller virial masses.
The HG virial masses are in excellent agreement with
the corresponding true masses.  This is to be expected, since the
massive halos which we broke up consisted originally of virialized clumps of
particles.  The NW algorithm, on the other hand, does a poor job
estimating group masses because its velocity linking length is too
small.

The Rm group masses, both true and virial, are not sensitive
to the breakup procedure.  Breakup Rm groups have smaller projected
and harmonic radii and larger velocity dispersions than do
raw groups, but these effect cancel each other out in the calculation
of the virial mass.

Groups identified in an absolute magnitude limited halo sample
using a fixed linking length (RM) have significantly different properties
than the groups found in the apparent magnitude limited halo catalogs
(Vm and Rm).  Statistics for the RM groups are given in
Tables 9 and 10.  These groups were identified in halo catalogs
with an absolute magnitude limit $M_{B(0)} < -16$; this is equivalent
to volume limiting out to 1000 km s$^{-1}$.  The linking length used
was $0.55h^{-1}$ Mpc, but the results are similar for a linking length
of $0.27h^{-1}$ Mpc, which is equal to the HG transverse
linking length $D_L$ at 1000 km s$^{-1}$.  The differences seen between
the raw and breakup Vm and Rm catalogs are also present in the RM
catalogs; {e.g.}, the fact that breakup groups are more massive than
raw groups.  But the distribution of RM group masses is significantly lower
than
those of the
Vm and RM masses, due to the fact that so many more faint halos
exist in the RM catalog.

Unfortunately, some and perhaps all of the differences between the
RM and Vm or Rm catalogs are due to the effect of the apparent magnitude limit
and hence are sensitive to the method by which our simulated halos
were illuminated.
The reader may note that the virial mass to light ratios in
raw RM groups are high compared to Vm or Rm groups; perhaps
this means that we underestimate the true $M_{\rm VT}/L$
of groups in the real data.  We cannot make this claim,
however, because of the manner in which our halos were illuminated.
The simulation produces many more faint halos than are observed
if we assume a constant mass to light ratio for illuminating
all halos.  Instead, we forced our luminosity function to fit
that of the CfA data.  As a result, the luminosity function of
halos rises more slowly than the mass function at the low end.
Consequently, our low mass halos were assigned even lower luminosities,
resulting in a high $M/L$ for faint halos.  This effect
is illustrated in Figure \ref{iii.3paper}, which shows $M_{\rm halo}/L$
as a function of $M_{\rm halo}$.  The spread in the relation is
because assigned luminosities are rank ordered by circular velocity,
which correlates well but not exactly with $M_{\rm halo}$.  The striped
effect arises from the quantization of true halo masses in units of the
particle mass.
Since our low mass, faint halos have high mass to light ratios,
our faint groups present in the RM catalogs will can be expected
to have high mass to light ratios also.

\subsection{Internal Properties of CfA Groups}
Although it cannot help us judge the accuracy of the group
finding algorithms, it is nonetheless interesting to look at
the properties of the HG and NW groups identified in the CfA
data.  These data are given in Tables 11 and 12.
Again we see that the NW groups have lower velocity dispersions and
virial masses than the HG groups.  The distributions of the three
independent group properties $r_h$, $\sigma_v$ and $M_{\rm VT}/L$ are all
narrower (as measured by the difference between the third and first
quartiles of the distribution), indicating that NW groups are a less diverse
class of objects than are HG groups.  Again, this comes at the expense
of biased velocity dispersions.  In addition to having smaller virial
masses, NW groups in the CfA data tend to be smaller in spatial
extent than HG groups.

NW groups in the CfA data we analyze have extremely low values of $M_{\rm
VT}/L$.
NW87 find a median mass to light ratio of $220h$ in the first CfA
survey, whereas we find a value of only $110h$ in the CfA extension.
The discrepancy here is probably due to the relative depths
of the original CfA survey and its extension, which is one magnitude deeper.
Almost all of the groups found by NW87 were less distant than 8000 km
s$^{-1}$,
which is about where the HG and NW scalings for the velocity linking
parameter begin to diverge rapidly.  As a result, the groups found by
NW87 in their more shallow galaxy sample were not significantly
affected by the biased velocity dispersions we see here.

\section{Group-Group Correlation Functions}
The correlation function $\xi(s)$ is another diagnostic by which
we can compare the NW and HG algorithms, and at the same time
study the effects of peculiar velocities on the group correlation
function by comparing Vm groups to Rm and RM groups.
We estimate the correlation function as
\begin{equation}
\xi(s)={n_R \over n_D}{N_{DD}(s) \over N_{DR}(s)} -1 \,,
\label{xis}
\end{equation}
\begin{equation}
s={(V_i^2 + V_j^2 -2 V_i V_j \cos \theta_{ij})^{1/2} \over H_0} \,,
\end{equation}
where $V_i$ and $V_j$ are the radial velocities of two galaxies with angular
separation $\theta_{ij}$ and $H_0$ is Hubble's constant.
In order to account for edge effects, a catalog of randomly distributed
points with geometry and selection function identical
to the real or simulated data must
be generated.  Then $n_D$ and $n_R$ are the number of points in the data
and the random catalog, respectively, $N_{DD}(s)$ is the number of pairs
in the data separated by redshift distance $s$, and $N_{DR}(s)$ is the number
of pairs, one from the data and one from the random catalog, separated by $s$.
We denote the galaxy-galaxy luminosity function as $\xi_{\rm gg}$ and
the group-group function as $\xi_{\rm GG}$.

Calculating the correlation functions requires constructing
catalogs of randomly distributed points in the same volume as
the real or simulated data and with the same selection function.
We use the same random catalogs for computing both $\xi_{gg}$ and
$\xi_{GG}$.  A comparison of the redshift distributions of simulated groups
to that of simulated galaxies revealed good agreement, justifying
our implicit assumption of the similarity of the simulated galaxy and group
selection functions.
In order to minimize
the statistical noise from the random catalog, we computed $\xi(s)$ for
ten separate simulated catalogs corresponding to our ten different observers,
using a different random catalog for each observer.  These results were then
averaged.

We were
also concerned with the effects the edges of the sample volume
might have on the group distribution, since small groups straddling
the edge of the volume may not be identified.  In fact, the simulated group
distribution does cut off abruptly at about $0.1^\circ$ from the
edges of the sample volume.  We compensated for this in our random
catalogs by masking out all points within $0.1^\circ$ of the
edges of the sample when computing $\xi_{\rm GG}$, and found this to have
almost no effect
on the resulting $\xi_{\rm GG}$ correlation functions.

Figure \ref{iii.1x} shows $\xi(s)$ for halos and for groups identified
in different ways from the raw catalogs.  The corresponding figure
for the breakup case is Figure \ref{iii.1.breakup}.
The general relationships between the different $\xi(s)$ curves
for the breakup catalogs and the raw catalogs are quite similar.  However,
as the correlation function of the breakup halos is higher than for the raw
case, the different $\xi(s)$ curves are squeezed together in Figure
\ref{iii.1.breakup}.

In both the apparent and absolute magnitude
limited samples, the group-group correlation function is
higher than that of the galaxies, as one would expect if clustering
is hierarchical (Bahcall \& Soneira 1983; Kaiser 1984; Bahcall 1988).
Ramella {et al.} (1990) find this not
to be the case in their analysis of three $6^\circ$ slices of the
CfA survey extension.  Unfortunately, the single slice of that same data
which we are studying does not
contain enough groups from which to construct a reliable correlation function.

For the simulated raw and breakup catalogs, the correlation functions
have similar slopes but differ in their amplitudes.
First note the behavior of
$\xi_{\rm gg}$, which is the main point of difference between
the raw halos and the breakup halos.
$\xi_{\rm gg}$ for the apparent magnitude limited (Vm) raw
catalog lies between the $\xi_{\rm gg}$ curves computed from raw catalogs
which were absolute magnitude limited at $M_{B(0)} \leq -16$
and $M_{B(0)} \leq -20$.  $M_{B(0)} = -16$ is the faintest halo
visible at 1000 km s$^{-1}$, while $M_{B(0)} = -20$ is the faintest halo
visible at 6500 km s$^{-1}$, which is the median redshift for our groups.
This behavior occurs because the correlation amplitude of our
simulated halos increases with luminosity.  This property
has been found in the original CfA survey (Hamilton 1988).
As one looks to more distant galaxies (or halos), one sees brighter
objects with intrinsically stronger clustering.  The correlation
function measured for an apparent magnitude limited sample like our Vm
catalog is a weighted average of the correlation functions of
objects of different luminosities.
This does not explain the fact that for the breakup groups, the Vm halos
are not any more correlated than the RM halos which have been
absolute magnitude limited at $-16$.  The reason for this is that
the correlation functions are squeezed together in the breakup
case until their error bars overlap.
In either case, the correlation amplitude
of the halos in the apparent magnitude limited catalog is larger
than that of the halos in the absolute magnitude limited catalog
by about 0.1 dex, or 25\%.

We find the same amplification of the correlation length
in the Vm groups relative to the RM groups.  Again, this is because
only groups containing bright and relatively strongly clustered
halos appear in much of the volume of the Vm and Rm samples.
The amount of the increase in the correlation length appears to be about
the same, or possible a little larger, for the halos than for the groups.

Because there are fewer NW than HG groups, the correlation functions
for the NW groups are noisier than but otherwise indistinguishable
from those of the HG groups.  This indicates that the two
algorithms are equally suited to determining the group-group
correlation function.  We noted above that the similarity
of the distributions of true mass between the HG and NW groups also
indicated that they tend to locate the same objects, although
the NW algorithm results in biased virial mass determinations.

Although the Rm groups appear to be more strongly correlated than the
Vm groups, these curves are noisy and their error bars (not shown) do overlap.
This behavior makes sense when one considers that the Vm groups typically
contain an Rm group plus some interlopers.  Then the spatial distribution
of Vm groups is the same as that of the Rm groups with a few spurious
groups added at uncorrelated positions.
Therefore it appears that the use of velocities rather than true
distances in identifying groups is adequate for the computation of a
redshift space group-group correlation function.

\section{Conclusions}
Comparing the internal properties of groups
identified by the HG and NW methods reveals a significant bias
in the NW groups in deep surveys toward low velocity dispersions and virial
masses.
This bias was not significant in the original application of the NW algorithm
(NW87) to the original CfA survey due to the relatively shallow depth of that
survey.
Nolthenius (1993) reports that in fact, a slight negative trend of
$M/L$ with distance was seen in NW87, as we would expect, and that the effect
was more
pronounced in his flow-model corrected group catalog.
This effect is simply interpreted as the result of group truncation
due to an overly restrictive velocity linking criterion, whereby
masses of groups at large redshift are systematically underestimated.
It is seen here in both
our simulated catalogs and in the real data from the CfA survey extension.
For the HG algorithm, on the other hand,
individual mass estimates for the breakup groups are unbiased, with the median
$\log_{10}(M_{\rm VT}/M_{\rm true}) = 0.003 \pm 0.075$.
The virial masses of raw HG groups are moderately overestimated,
with the median $M_{\rm VT}$ about 50\% larger than the median $M_{\rm true}$,
though the error bars on each of these median values is almost as large
as the difference between them.
Because our breakup galaxy catalogs match the clustering of real
(CfA) galaxies better than do our raw catalogs, we expect the accuracy
of virial masses for real groups to be close to that seen here for
simulated breakup groups when the HG algorithm is applied.

Diaferio et al. (1993) find that the spread in group masses found
by RGH89 is consistent with the spread in virial mass estimates
of a single collapsing loose group viewed from different angles.
We find that the distribution of virial masses is broader than that
of true masses for both our raw and breakup HG groups.
Projection effects may be responsible for this spread.

We find that the NW and the HG algorithms produce groups with similar
correlation functions $\xi_{\rm GG}(s)$, although the HG algorithm is
slightly better suited to this task because it produces more groups, for better
statistics.  Comparing $\xi_{\rm GG}$ computed from our Vm and Rm
groups indicates that the use of velocities rather than true distances
for identifying groups
does not affect group clustering as measured by the correlation function
$\xi_{GG}(s)$.

The virial masses of our simulated groups are higher for our breakup groups
than for raw groups, and in both cases are much higher than
the virial masses of CfA groups.
This may be an important shortcoming of the CDM model,
as this is a purely dynamical effect, not one which can be
ascribed to our halo illumination technique or to the lack of gas
dynamics in the simulation.  Our CDM groups are also more strongly
correlated than individual CDM halos, contrary to the finding of
Ramella et al. (1990) for groups found in the first two slices
of the CfA redshift survey extension.

Properties of our RM groups differ significantly from those of our Vm and Rm
groups.  This fact should serve as a reminder to those performing
cosmological structure simulations that it is important to compare
simulated data to observations as accurately as possible, by mimicking
the observation procedure and including important effects such as a flux
limit.

\acknowledgments
Supercomputer time was provided by the Cornell National Supercomputer
Facility and the National Center for Supercomputing Applications.
Support was provided by NSF grants AST90-01762 and ASC93-18185
and by an NSF graduate fellowship.
The author is grateful for the guidance of Ed Bertschinger
and the previous work of Bertschinger and James Gelb.

\section{Appendix:  Definitions of Group Properties}
The quantities in Tables 1 through 12 are defined as follows:
The number of members in a group is $N$.
The true mass measured from the simulation is denoted $M_{\rm true}$
and is computed from the largest number of those particles grouped
together by DENMAX
which are mutually gravitationally bound.
The group mass quoted is the sum of the masses
of the halos in all RM groups whose membership totally or
partially overlaps the membership of the Vm or Rm group.

A statistical correction is applied to group luminosities to account
for group members which are below the magnitude limit of the catalog:
\begin{equation}
L=L_0 {\int_0^\infty L \phi(L) dL \over \int_{L_V}^\infty L \phi(L) dL} \,,
\end{equation}
where $L_0$ is the sum of the luminosities of the visible members and
$L_v$ is the minimum luminosity visible at a redshift $V$.
When computing the mass to light ratio $M_{\rm true}/L$
we use true distances of individual group members to convert
their apparent magnitudes to luminosities and the true distance of the group
center to make the group luminosity correction.  For the $M_{\rm VT}/L$
statistic we used redshift distances for computing both individual
member luminosities and the group luminosity correction.  We also
tested using the group's mean redshift distance for computing
individual members' luminosities from their magnitudes.  This made no
significant difference in the median group mass to light ratio.

The remaining quantities are defined as by NW87.
The mean harmonic radius, $r_h$, is given by
\begin{equation}
r_h={\pi \over 2} {V \over H_0} N(N-1) \bigg[\sum_{i<j} {1 \over \tan
\theta_{ij} / 2} \bigg]^{-1} \,.
\end{equation}
The dispersions in both velocity and true distance, $\sigma_v$ and $\sigma_r$,
are
\begin{equation}
\sigma_s = \bigg[{1 \over N-1} \sum_{i=1}^N (s_i - \langle s \rangle )^2
\bigg]^{1/2},
\end{equation}
where $s$ can refer to either redshift ($v$) or true ($r$) distance and
$\langle s \rangle$ is the arithmetic mean for the group.
The virial crossing time, in units of the Hubble time, is
\begin{equation}
H_0 t_c={2 \over \sqrt{3}} {r_h H_0 \over \sigma_v} \,.
\end{equation}
This measure of the crossing time is equal to 0.343 times the collapse
time of a
growing homogeneous spherical perturbation.
Groups with collapse times less than the Hubble time will have
crossing times less than $0.343H_0^{-1}$.  Groups with longer
crossing times cannot be expected to be virialized.

Virial mass is estimated as
\begin{equation}
M_{\rm VT}={6 \sigma_v^2 r_h \over G} \,.
\end{equation}

\clearpage

\begin{table}
\centerline{Table 1: Internal Properties of Raw HG Vm Groups$^*$}
\begin{center}
\vspace{00 pt}
{\small
\begin{tabular}{lrrrr}
 & Mean & 1st Quartile & Median & 3rd Quartile\\
\hline
$N$&$   3.80 \pm 0.22$&$3.00 \pm 0.00$&$3.00 \pm 0.00$&$4.10 \pm 0.30$\\
$\log_{10}(M_{\rm true}/M_{\odot})$&$13.54 \pm 0.25$&$13.40 \pm 0.15$&$13.82
\pm 0.17$&$14.20 \pm 0.14$\\
$\log_{10}(M_{\rm VT}/M_{\odot})$&$13.91 \pm 0.12$&$13.45 \pm 0.14$&$13.98 \pm
0.14$&$14.41 \pm 0.17$\\
$\log_{10}(M_{\rm VT}/M_{\rm true})$&$0.366 \pm 0.262$&$-0.224 \pm
0.067$&$0.159 \pm 0.121$&$0.620 \pm 0.122$\\
$r_h$ (Mpc)&$1.23 \pm 0.14$&$0.81 \pm 0.12$&$1.14 \pm 0.15$&$1.57 \pm 0.20$\\
$\sigma_v$ (km s$^{-1}$)&$291.3 \pm 25.5$&$154.8 \pm 28.5$&$259.7 \pm
22.3$&$391.5 \pm 47.0$\\
$H_0 \sigma_r$ (km s$^{-1}$)&$242.2 \pm 57.1$&$51.7 \pm 19.8$&$155.7 \pm
61.7$&$370.4 \pm 108.9$\\
$H_0t_c$&$0.431 \pm 0.119$&$0.159 \pm 0.020$&$0.247 \pm 0.017$&$0.435 \pm
0.057$\\
$M_{\rm true}/L~~(M_\odot / L_\odot)$&$346.0 \pm 202.7$&$142.2 \pm 10.2$&$207.4
\pm 18.9$&$328.7 \pm 56.7$\\
$M_{\rm VT}/L~~(M_\odot / L_\odot)$&$1041.6 \pm 1450.8$&$135.9 \pm 34.6$&$369.0
\pm 84.2$&$773.3 \pm  211.9$\\
\hline
\hline
\end{tabular}
}
\end{center}
$^*$~Dependences on $h$ have been included in the quoted values for this
and subsequent tables.  We use $h=0.5$.
\end{table}

\begin{table}
\centerline{Table 2: Internal Properties of Breakup HG Vm Groups}
\begin{center}
\vspace{12 pt}
{\small
\begin{tabular}{lrrrr}
 & Mean & 1st Quartile & Median & 3rd Quartile\\
\hline
$N$&$6.43 \pm 1.01$&$3.00 \pm  0.00$&$4.50 \pm 0.67$&$7.20 \pm 1.29$\\
$\log_{10}(M_{\rm true}/M_{\odot})$&$13.98 \pm 0.11$&$13.70 \pm 0.10$&$14.02
\pm 0.11$&$14.37 \pm 0.08$\\
$\log_{10}(M_{\rm VT}/M_{\odot})$&$13.99 \pm 0.15$&$13.52 \pm 0.16$&$14.05 \pm
0.18$&$14.53 \pm 0.17$\\
$\log_{10}(M_{\rm VT}/M_{\rm true})$&$0.011 \pm 0.120$&$-0.360 \pm
0.084$&$0.003 \pm 0.075$&$0.318 \pm 0.098$\\
$r_h$ (Mpc)&$0.95 \pm 0.12$&$0.53 \pm 0.08$&$0.80 \pm 0.08$&$1.24 \pm 0.23$\\
$\sigma_v$ (km s$^{-1}$)&$368.1 \pm 39.4$&$204.7 \pm 33.8$&$329.4 \pm
29.3$&$497.4 \pm 57.6$\\
$H_0 \sigma_r$ (km s$^{-1}$)&$164.4 \pm 45.6$&$14.9 \pm 2.8$&$42.5 \pm
14.3$&$225.8 \pm 103.22$\\
$H_0t_c$&$0.217 \pm 0.039$&$0.090 \pm 0.007$&$0.143 \pm 0.020$&$0.246 \pm
0.031$\\
$M_{\rm true}/L~~(M_\odot / L_\odot)$&$456.9 \pm 87.9$&$251.1 \pm 35.2$&$347.9
\pm 27.0$&$454.6 \pm 47.8$\\
$M_{\rm VT}/L~~(M_\odot / L_\odot)$&$525.6 \pm 115.2$&$153.5 \pm 27.9$&$334.0
\pm 43.1$&$666.3 \pm 118.5$\\
\hline
\hline
\end{tabular}
}
\end{center}
\end{table}

\clearpage

\begin{table}
\centerline{Table 3: Internal Properties of Raw NW Vm Groups}
\begin{center}
\vspace{12 pt}
{\small
\begin{tabular}{lrrrr}
 & Mean & 1st Quartile & Median & 3rd Quartile\\
\hline
$N$&$3.42 \pm 0.12$&$3.00 \pm 0.00$&$3.00 \pm 0.00$&$3.90 \pm 0.30$\\
$log_{10}(M_{\rm true}/M_{\odot})$&$13.65 \pm  0.27$&$13.35 \pm  0.25$&$13.82
\pm  0.22$&$14.19 \pm  0.19$\\
$\log_{10}(M_{\rm VT}/M_{\odot})$&$13.47 \pm 0.15$&$13.09 \pm 0.18$&$13.53 \pm
0.17$&$13.93 \pm 0.16$\\
$\log_{10}(M_{\rm VT}/M_{\rm true})$&$  -0.183 \pm    0.230$&$  -0.651 \pm
0.132$&$  -0.240 \pm    0.085$&$   0.095 \pm    0.084$\\
$r_h$ (Mpc)&$1.18 \pm 0.15$&$0.71 \pm 0.10$&$1.06 \pm 0.14$&$1.45 \pm 0.20$\\
$\sigma_v$ (km s$^{-1}$)&$170.9 \pm 18.8$&$99.9 \pm 12.8$&$160.9 \pm
19.7$&$234.1 \pm 31.0$\\
$H_0 \sigma_r$ (km s$^{-1}$)&$150.8 \pm 46.0$&$35.8 \pm 10.0$&$89.2 \pm
25.8$&$253.8 \pm 88.9$\\
$H_0t_c$&$0.560 \pm 0.067$&$0.245 \pm 0.033$&$0.391 \pm 0.036$&$0.682 \pm
0.105$\\
$M_{\rm true}/L~~(M_\odot / L_\odot)$&$ 303.7 \pm   77.7$&$ 142.0 \pm   15.3$&$
211.1 \pm   29.2$&$ 336.9 \pm   55.8$\\
$M_{\rm VT}/L~~(M_\odot / L_\odot)$&$1468.3 \pm 2571.7$&$47.9 \pm 17.2$&$112.0
\pm 24.6$&$278.3 \pm 81.4$\\
\hline
\hline
\end{tabular}
}
\end{center}
\end{table}

\begin{table}
\centerline{Table 4: Internal Properties of Breakup NW Vm Groups}
\begin{center}
\vspace{12 pt}
{\small
\begin{tabular}{lrrrr}
 & Mean & 1st Quartile & Median & 3rd Quartile\\
\hline
$N$&$5.14 \pm 0.55$&$3.00 \pm 0.00$&$3.80 \pm 0.40$&$5.53 \pm 0.63$\\
$\log_{10}(M_{\rm true}/M_{\odot})$&$13.98 \pm  0.16$&$13.74 \pm  0.11$&$14.05
\pm  0.11$&$14.39 \pm  0.10$\\
$\log_{10}(M_{\rm VT}/M_{\odot})$&$13.40 \pm  0.09$&$13.04 \pm  0.09$&$13.42
\pm  0.06$&$13.82 \pm  0.12$\\
$\log_{10}(M_{\rm VT}/M_{\rm true})$&$-0.575 \pm  0.122$&$-0.965 \pm
0.098$&$-0.587 \pm  0.038$&$-0.257 \pm  0.065$\\
$r_h$ (Mpc)&$0.87 \pm 0.10$&$0.45 \pm 0.05$&$0.68 \pm 0.07$&$1.04 \pm 0.11$\\
$\sigma_v$ (km s$^{-1}$)&$187.3 \pm  13.8$&$118.5 \pm  15.1$&$171.1 \pm
10.4$&$238.1 \pm  16.8$\\
$H_0 \sigma_r$ (km s$^{-1}$)&$ 81.9 \pm  15.2$&$ 11.4 \pm   2.0$&$ 21.8 \pm
2.5$&$ 66.6 \pm  19.3$\\
$H_0t_c$&$0.361 \pm 0.081$&$0.143 \pm 0.020$&$0.230 \pm 0.030$&$0.423 \pm
0.077$\\
$M_{\rm true}/L~~(M_\odot / L_\odot)$&$ 786.6 \pm  211.0$&$ 364.3 \pm   40.2$&$
518.4 \pm   59.3$&$ 827.0 \pm  164.6$\\
$M_{\rm VT}/L~~(M_\odot / L_\odot)$&$ 147.1 \pm   81.3$&$  57.3 \pm   15.1$&$
119.8 \pm   25.4$&$ 249.6 \pm   43.1$\\
\hline
\hline
\end{tabular}
}
\end{center}
\end{table}

\clearpage

\begin{table}
\centerline{Table 5: Internal Properties of Raw HG Rm Groups}
\begin{center}
\vspace{12 pt}
{\small
\begin{tabular}{lrrrr}
 & Mean & 1st Quartile & Median & 3rd Quartile\\
\hline
$N$&$3.40 \pm 0.23$&$3.00 \pm 0.00$&$3.00 \pm 0.00$&$3.55 \pm 0.43$\\
$log_{10}(M_{\rm true}/M_{\odot})$&$13.93 \pm  0.21$&$13.53 \pm  0.29$&$14.00
\pm  0.24$&$14.34 \pm  0.22$\\
$log_{10}(M_{\rm VT}/M_{\odot})$&$13.92 \pm  0.14$&$13.43 \pm  0.20$&$14.00 \pm
 0.17$&$14.51 \pm  0.13$\\
$log_{10}(M_{\rm VT}/M_{\rm true})$&$-0.006 \pm  0.189$&$-0.353 \pm  0.200$&$
0.036 \pm  0.132$&$ 0.406 \pm  0.145$\\
$r_h$ (Mpc)&$1.16 \pm 0.17$&$0.80 \pm 0.13$&$1.10 \pm 0.18$&$1.51 \pm 0.28$\\
$\sigma_v$ (km s$^{-1}$)&$319.7 \pm  39.0$&$151.8 \pm  32.7$&$276.7 \pm
33.7$&$437.9 \pm  66.0$\\
$H_0 \sigma_r$ (km s$^{-1}$)&$ 25.3 \pm   4.0$&$ 15.0 \pm   4.9$&$ 24.7 \pm
4.0$&$ 33.4 \pm   5.3$\\
$H_0t_c$&$0.401 \pm 0.120$&$0.147 \pm 0.044$&$0.226 \pm 0.049$&$0.402 \pm
0.081$\\
$M_{\rm true}/L~~(M_\odot / L_\odot)$&$ 347.2 \pm  119.0$&$ 187.8 \pm   55.8$&$
267.2 \pm   74.2$&$ 364.5 \pm   87.6$\\
$M_{\rm VT}/L~~(M_\odot / L_\odot)$&$ 632.1 \pm  463.9$&$ 113.3 \pm   61.2$&$
276.9 \pm  132.6$&$ 857.7 \pm  705.2$\\
\hline
\hline
\end{tabular}
}
\end{center}
\end{table}

\begin{table}
\centerline{Table 6: Internal Properties of Breakup HG Rm Groups}
\begin{center}
\vspace{12 pt}
{\small
\begin{tabular}{lrrrr}
 & Mean & 1st Quartile & Median & 3rd Quartile\\
\hline
$N$&$6.91 \pm 1.36$&$3.30 \pm 0.46$&$4.90 \pm 0.94$&$7.68 \pm 1.69$\\
$log_{10}(M_{\rm true}/M_{\odot})$&$14.02 \pm  0.07$&$13.71 \pm  0.08$&$13.99
\pm  0.09$&$14.36 \pm  0.09$\\
$log_{10}(M_{\rm VT}/M_{\odot})$&$13.95 \pm  0.09$&$13.55 \pm  0.10$&$14.00 \pm
 0.10$&$14.47 \pm  0.10$\\
$log_{10}(M_{\rm VT}/M_{\rm true})$&$-0.066 \pm  0.066$&$-0.324 \pm  0.084$&$
0.019 \pm  0.074$&$ 0.263 \pm  0.072$\\
$r_h$ (Mpc)&$0.75 \pm 0.11$&$0.43 \pm 0.05$&$0.67 \pm 0.07$&$0.99 \pm 0.15$\\
$\sigma_v$ (km s$^{-1}$)&$385.0 \pm  30.4$&$218.1 \pm  25.1$&$339.5 \pm
29.3$&$526.2 \pm  45.7$\\
$H_0 \sigma_r$ (km s$^{-1}$)&$ 19.2 \pm   1.9$&$ 10.7 \pm   1.9$&$ 17.3 \pm
1.7$&$ 25.3 \pm   2.5$\\
$H_0t_c$&$0.181 \pm 0.058$&$0.066 \pm 0.009$&$0.112 \pm 0.021$&$0.193 \pm
0.044$\\
$M_{\rm true}/L~~(M_\odot / L_\odot)$&$ 427.2 \pm   80.4$&$ 289.3 \pm   17.7$&$
354.3 \pm   10.9$&$ 432.2 \pm   34.5$\\
$M_{\rm VT}/L~~(M_\odot / L_\odot)$&$ 490.4 \pm   82.0$&$ 164.4 \pm   59.7$&$
335.1 \pm   69.9$&$ 664.8 \pm  141.8$\\
\hline
\hline
\end{tabular}
}
\end{center}
\end{table}

\clearpage

\begin{table}
\centerline{Table 7: Internal Properties of Raw NW Rm Groups}
\begin{center}
\vspace{12 pt}
{\small
\begin{tabular}{lrrrr}
 & Mean & 1st Quartile & Median & 3rd Quartile\\
\hline
$N$&$3.29 \pm 0.23$&$3.00 \pm 0.00$&$3.00 \pm 0.00$&$3.10 \pm 0.30$\\
$log_{10}(M_{\rm true}/M_{\odot})$&$14.02 \pm  0.18$&$13.69 \pm  0.35$&$14.08
\pm  0.24$&$14.42 \pm  0.13$\\
$log_{10}(M_{\rm VT}/M_{\odot})$&$14.01 \pm  0.16$&$13.54 \pm  0.25$&$14.09 \pm
 0.24$&$14.58 \pm  0.16$\\
$log_{10}(M_{\rm VT}/M_{\rm true})$&$-0.011 \pm  0.171$&$-0.309 \pm  0.153$&$
0.039 \pm  0.141$&$ 0.346 \pm  0.173$\\
$r_h$ (Mpc)&$1.16 \pm 0.15$&$0.75 \pm 0.12$&$1.03 \pm 0.18$&$1.45 \pm 0.21$\\
$\sigma_v$ (km s$^{-1}$)&$353.3 \pm  48.3$&$181.1 \pm  63.5$&$305.3 \pm
60.7$&$475.9 \pm  83.8$\\
$H_0 \sigma_r$ (km s$^{-1}$)&$ 26.3 \pm   3.8$&$ 14.3 \pm   3.8$&$ 24.8 \pm
3.7$&$ 35.1 \pm   6.8$\\
$H_0t_c$&$0.359 \pm 0.102$&$0.126 \pm 0.039$&$0.198 \pm 0.038$&$0.360 \pm
0.107$\\
$M_{\rm true}/L~~(M_\odot / L_\odot)$&$ 276.5 \pm   84.2$&$ 143.4 \pm   24.0$&$
215.8 \pm   52.8$&$ 332.8 \pm   61.0$\\
$M_{\rm VT}/L~~(M_\odot / L_\odot)$&$ 599.5 \pm  482.3$&$ 104.4 \pm   77.6$&$
218.9 \pm  127.2$&$ 843.2 \pm  754.6$\\
\hline
\hline
\end{tabular}
}
\end{center}
\end{table}

\begin{table}
\centerline{Table 8: Internal Properties of Breakup NW Rm Groups}
\begin{center}
\vspace{12 pt}
{\small
\begin{tabular}{lrrrr}
 & Mean & 1st Quartile & Median & 3rd Quartile\\
\hline
$N$&$6.70 \pm 1.14$&$3.12 \pm 0.30$&$4.55 \pm 0.85$&$7.60 \pm 1.43$\\
$log_{10}(M_{\rm true}/M_{\odot})$&$13.99 \pm  0.12$&$13.70 \pm  0.10$&$14.00
\pm  0.10$&$14.37 \pm  0.10$\\
$log_{10}(M_{\rm VT}/M_{\odot})$&$13.93 \pm  0.09$&$13.51 \pm  0.13$&$13.97 \pm
 0.09$&$14.44 \pm  0.13$\\
$log_{10}(M_{\rm VT}/M_{\rm true})$&$-0.064 \pm  0.117$&$-0.376 \pm  0.077$&$
0.003 \pm  0.069$&$ 0.237 \pm  0.089$\\
$r_h$ (Mpc)&$0.71 \pm 0.11$&$0.42 \pm 0.04$&$0.63 \pm 0.07$&$0.89 \pm 0.14$\\
$\sigma_v$ (km s$^{-1}$)&$387.2 \pm  32.1$&$215.4 \pm  23.7$&$341.7 \pm
24.1$&$528.1 \pm  46.1$\\
$H_0 \sigma_r$ (km s$^{-1}$)&$ 17.9 \pm   1.8$&$ 10.2 \pm   1.4$&$ 16.0 \pm
1.5$&$ 23.5 \pm   2.8$\\
$H_0t_c$&$0.179 \pm 0.075$&$0.062 \pm 0.010$&$0.105 \pm 0.017$&$0.183 \pm
0.049$\\
$M_{\rm true}/L~~(M_\odot / L_\odot)$&$ 460.1 \pm   99.3$&$ 308.9 \pm   17.0$&$
366.5 \pm   21.4$&$ 461.1 \pm   37.8$\\
$M_{\rm VT}/L~~(M_\odot / L_\odot)$&$ 435.3 \pm  172.5$&$ 164.2 \pm   64.0$&$
334.1 \pm   70.1$&$ 658.2 \pm  140.9$\\
\hline
\hline
\end{tabular}
}
\end{center}
\end{table}

\clearpage

\begin{table}
\centerline{Table 9: Internal Properties of Raw RM Groups}
\begin{center}
\vspace{12 pt}
{\small
\begin{tabular}{lrrrr}
 & Mean & 1st Quartile & Median & 3rd Quartile\\
\hline
$N$&$3.90 \pm 0.05$&$3.00 \pm 0.00$&$3.00 \pm 0.00$&$4.00 \pm 0.00$\\
$log_{10}(M_{\rm true}/M_{\odot})$&$12.70 \pm  0.02$&$12.25 \pm  0.02$&$12.60
\pm  0.02$&$13.11 \pm  0.03$\\
$log_{10}(M_{\rm VT}/M_{\odot})$&$12.95 \pm  0.05$&$12.38 \pm  0.05$&$12.99 \pm
 0.04$&$13.56 \pm  0.06$\\
$log_{10}(M_{\rm VT}/M_{\rm true})$&$ 0.245 \pm  0.038$&$-0.225 \pm  0.022$&$
0.209 \pm  0.024$&$ 0.686 \pm  0.047$\\
$r_h$ (Mpc)&$0.98 \pm 0.03$&$0.72 \pm 0.02$&$0.97 \pm 0.03$&$1.23 \pm 0.04$\\
$\sigma_v$ (km s$^{-1}$)&$125.2 \pm   5.9$&$ 45.0 \pm   1.8$&$ 89.3 \pm
3.7$&$162.4 \pm   8.8$\\
$H_0 \sigma_r$ (km s$^{-1}$)&$ 19.2 \pm   0.1$&$ 12.7 \pm   0.1$&$ 18.3 \pm
0.2$&$ 24.8 \pm   0.3$\\
$H_0t_c$&$1.095 \pm 0.099$&$0.317 \pm 0.011$&$0.590 \pm 0.024$&$1.144 \pm
0.026$\\
$M_{\rm true}/L~~(M_\odot / L_\odot)$&$ 289.6 \pm    2.5$&$ 193.0 \pm    1.7$&$
229.1 \pm    2.9$&$ 315.0 \pm    6.1$\\
$M_{\rm VT}/L~~(M_\odot / L_\odot)$&$2001.6 \pm  265.4$&$ 154.4 \pm    8.5$&$
456.8 \pm   20.8$&$1337.6 \pm  109.5$\\
\hline
\hline
\end{tabular}
}
\end{center}
\end{table}

\begin{table}
\centerline{Table 10: Internal Properties of Breakup RM Groups}
\begin{center}
\vspace{12 pt}
{\small
\begin{tabular}{lrrrr}
 & Mean & 1st Quartile & Median & 3rd Quartile\\
\hline
$N$&$ 24.69 \pm   1.10$&$  3.00 \pm   0.00$&$ 14.00 \pm   0.89$&$ 29.95 \pm
2.10$\\
$log_{10}(M_{\rm true}/M_{\odot})$&$13.35 \pm  0.02$&$13.04 \pm  0.06$&$13.36
\pm  0.02$&$13.68 \pm  0.02$\\
$log_{10}(M_{\rm VT}/M_{\odot})$&$13.41 \pm  0.03$&$13.12 \pm  0.03$&$13.48 \pm
 0.03$&$13.79 \pm  0.03$\\
$log_{10}(M_{\rm VT}/M_{\rm true})$&$ 0.052 \pm  0.018$&$-0.142 \pm  0.018$&$
0.020 \pm  0.015$&$ 0.193 \pm  0.017$\\
$r_h$ (Mpc)&$0.62 \pm 0.02$&$0.39 \pm 0.01$&$0.52 \pm 0.02$&$0.78 \pm 0.03$\\
$\sigma_v$ (km s$^{-1}$)&$232.9 \pm   5.5$&$134.3 \pm   7.9$&$222.1 \pm
6.0$&$299.0 \pm   6.2$\\
$H_0 \sigma_r$ (km s$^{-1}$)&$ 16.6 \pm   0.4$&$ 11.3 \pm   0.2$&$ 15.3 \pm
0.5$&$ 20.2 \pm   0.5$\\
$H_0t_c$&$0.354 \pm 0.040$&$0.082 \pm 0.004$&$0.120 \pm 0.003$&$0.325 \pm
0.037$\\
$M_{\rm true}/L~~(M_\odot / L_\odot)$&$ 492.0 \pm   44.6$&$ 318.9 \pm    2.3$&$
363.5 \pm    3.2$&$ 428.4 \pm    5.5$\\
$M_{\rm VT}/L~~(M_\odot / L_\odot)$&$1325.1 \pm  218.0$&$ 252.6 \pm   10.0$&$
362.2 \pm    9.1$&$ 622.8 \pm   30.6$\\
\hline
\hline
\end{tabular}
}
\end{center}
\end{table}

\clearpage

\begin{table}
\centerline{Table 11: Internal Properties of CfA HG Groups}
\begin{center}
\vspace{12 pt}
{\small
\begin{tabular}{lrrrr}
 & Mean & 1st Quartile & Median & 3rd Quartile\\
\hline
$N$& 6.67 & 3.00 & 4.00 & 6.00 \\
$\log_{10}(M_{\rm VT}/M_\odot)$& 13.70 & 12.93 & 13.71 & 14.47 \\
$r_h$ (Mpc)& 1.08 & 0.38 & 0.81 & 1.67 \\
$\sigma_v$ (km s$^{-1}$)& 290.8 & 121.5 & 215.3 & 398.3 \\
$H_0t_c$& 0.288 & 0.117 & 0.218 & 0.352 \\
$M_{\rm VT}/L~~(M_\odot / L_\odot)$& 3365 & 38 & 204 & 572 \\
\hline
\hline
\end{tabular}
}
\end{center}
\end{table}

\begin{table}
\centerline{Table 12: Internal Properties of CfA NW Groups}
\begin{center}
\vspace{12 pt}
{\small
\begin{tabular}{lrrrr}
 & Mean & 1st Quartile & Median & 3rd Quartile\\
\hline
$N$& 5.59 & 3.00 & 4.00 & 6.00 \\
$\log_{10}(M_{\rm VT}/M_\odot)$& 13.22 & 12.74 & 13.27 & 13.85 \\
$r_h$ (Mpc)& 0.85 & 0.35 & 0.64 & 1.13 \\
$\sigma_v$ (km s$^{-1}$)& 177.2 &  78.3 & 153.0 & 241.4 \\
$H_0t_c$& 0.404 & 0.155 & 0.250 & 0.461 \\
$M_{\rm VT}/L~~(M_\odot / L_\odot)$& 973 & 20 & 55 & 188 \\
\hline
\hline
\end{tabular}
}
\end{center}
\end{table}

\clearpage

\begin{figure}
    \caption{ $M_{\rm halo}/L$ vs $M_{\rm halo}$ for the halos in the
simulation. }
    \label{iii.3paper}
\end{figure}

\begin{figure}
    \caption{ Two point correlation functions for simulated halos and groups
in the raw catalogs,
each calculated for ten catalogs corresponding to different
observers and averaged over those ten catalogs.  Arrows on the $s$ axis
near 0.7 and 0.8 indicate the correlation lengths quoted by Ramella et al.
(1990) for galaxies and for groups, respectively.
A representative error bar for $\xi_{GG}$ is also plotted.  Error bars
for $\xi_{gg}$ are much smaller.
The symbol key gives absolute magnitude limit for the RM catalogs
and the value of the linking length used for the groups.  $D_0$ is
the transverse linking length for the Vm and Rm groups and the total
three dimensional linking length for the RM groups.
Correlation functions for NW groups, not shown here, are noisier but otherwise
indistinguishable from those of HG groups.}
    \label{iii.1x}
\end{figure}

\begin{figure}
    \caption{Same as Fig. \protect\ref{iii.1x}, but for the breakup catalogs.}
    \label{iii.1.breakup}
\end{figure}

\clearpage
\end{document}